\documentclass[pdflatex,sn-mathphys-num]{sn-jnl}

\usepackage{graphicx}%
\usepackage{multirow}%
\usepackage{amsmath,amssymb,amsfonts}%
\usepackage{amsthm}%
\usepackage[title]{appendix}%
\usepackage{xcolor}%
\usepackage{textcomp}%
\usepackage{manyfoot}%
\usepackage{booktabs}%
\usepackage{algorithm}%
\usepackage{algorithmicx}%
\usepackage{algpseudocode}%
\usepackage{listings}%
\usepackage{bookmark}
\usepackage{comment}
\usepackage{float}
\usepackage{placeins}
\usepackage{adjustbox}
\usepackage{array}
\usepackage{makecell}
\usepackage{tabularx}

\theoremstyle{thmstyleone}%

\theoremstyle{thmstyletwo}%

\theoremstyle{thmstylethree}%

\begin{document}

\title[Clustering-Augmented IPW]{Clustering-Informed Inverse Probability Weighting Strategies for Causal Effect Estimation in Observational Studies}

\author*[1,2]{\fnm{Ruohui} \sur{Chen}} \email{chenruohui2013@gmail.com}
\author[1]{\fnm{Scott} \sur{Zuo}} 
\author[3]{\fnm{Whitney} \sur{Stevens}} 
\author[3]{\fnm{Seth} \sur{Pollack}} 
\author[1]{\fnm{Wenna} \sur{Xi}} 
\author[1]{\fnm{Lucia} \sur{Petito}} 
\author[1]{\fnm{Lihui} \sur{Zhao}} 
\author[1]{\fnm{Hui} \sur{Zhang}} 
\affil*[1]{\orgdiv{Department of Preventive Medicine}, \orgname{Northwestern University}, \orgaddress{\city{Chicago}, \state{IL}, \country{USA}}}
\affil[2]{\orgdiv{Department of Biostatistics and Bioinformatics}, \orgname{Moffitt Cancer Center and Research Institute}, \orgaddress{\city{Tampa}, \state{FL}, \country{USA}}}
\affil[3]{\orgdiv{Feinberg School of Medicine}, \orgname{Northwestern University}, \orgaddress{\city{Chicago}, \state{IL}, \country{USA}}}

\abstract{Inverse probability weighting (IPW) is widely used to estimate causal effects in observational studies, but its performance depends on adequate propensity-score specification. Rather than proposing a single new estimator, we provide a rigorous head-to-head comparison of three propensity-score strategies for handling this kind of heterogeneity: standard IPW, a clustering-augmented IPW framework that fits separate propensity-score models within empirically derived clusters, and a global propensity-score model that instead includes estimated cluster membership as a covariate. Using simulations with and without embedded latent cluster structure, and under correctly specified and omitted-covariate propensity-score models, we evaluate each strategy's bias, mean squared error (MSE), and confidence-interval coverage across sample sizes ranging from 100 to 500. Both cluster-informed strategies substantially reduced the bias and MSE caused by omitted-covariate misspecification relative to standard IPW, but neither uniformly dominated the other: Clustering + IPW achieved lower MSE under correctly specified latent-cluster structure, while the global model with cluster labels generally achieved better coverage and lower bias at smaller sample sizes. We further apply the framework to 966 breast cancer patients treated with carboplatin, using generalized propensity scores to estimate the dose–response relationship between treatment cycles and hypersensitivity reaction risk; the clustered and standard analyses produced consistent pooled estimates, while the clustered analysis additionally yielded cluster-specific estimates and diagnostic profiles useful for clinical interpretation. These findings offer practical guidance for settings with suspected treatment-assignment heterogeneity: cluster-informed strategies may improve robustness to omitted-covariate propensity-score misspecification, while their relative performance depends on the underlying subgroup structure, sample size, and the relative importance of MSE, bias, and interval coverage.}


\maketitle

\section{Introduction}

Causal inference in observational studies often relies on inverse probability weighting (IPW) to adjust for confounding by reweighting individuals according to the probability of receiving the treatment of interest \cite{Seaman2013IPWreview,Zhou2021IPWreview2,Chang2022PSreview,Lunceford2004Stratification,Chen2024IPWDR}. The validity of IPW depends critically on adequate measurement of confounders and correct specification of the propensity-score model; in particular, misspecifying confounder--exposure associations can bias propensity-score-based effect estimates \cite{Schuster2023Misspecification}. IPW can also produce unstable weights and inflated variance when treatment assignment is highly imbalanced. These challenges are especially relevant in medical research, where relationships among treatment, covariates, and outcomes may be nonlinear, heterogeneous, or influenced by latent subgroups.

A key difficulty arises when treatment assignment and outcome risk vary across latent or poorly characterized patient subgroups, especially when this heterogeneity is only partially captured by observed pre-treatment covariates. Global propensity score models, fitted across the entire population, may fail to capture these latent structures, producing extreme weights or residual confounding. This motivates the use of unsupervised learning techniques, such as clustering, to identify more homogeneous subgroups that share similar baseline characteristics. By stratifying patients into latent subpopulations, clustering can reduce model dependence and improve balance between treated and untreated groups, offering a more data-driven foundation for causal inference \cite{Nagin2005GroupBased,Pan2013Cluster}.

Clustering methods, including hierarchical clustering (HC), K-means, and partitioning around medoids (PAM), have been successfully applied for subgroup discovery in biomedical studies \cite{Li2022HC,KaufmanPAM}. However, these methods are not directly designed for causal estimation and can fail to account for confounding, particularly if clusters do not align with treatment assignment patterns. When combined with IPW, clustering offers a hybrid strategy that models treatment assignment within smaller, more homogeneous strata, thereby reducing the risk of extreme weights and increasing robustness to misspecification.

Our work is closely related to, but distinct from, recent developments in causal subgroup analysis and localized propensity-score estimation. Existing subgroup propensity-score methods emphasize that covariate balance in the overall sample does not necessarily guarantee adequate balance within clinically meaningful subgroups. For example, Yang et al. developed propensity-score weighting methods for causal subgroup analysis and formalized subgroup-specific weighted average treatment effects, while Dong et al. proposed subgroup balancing propensity scores to address the bias--variance tradeoff that arises when attempting to improve covariate balance within subgroups \cite{Yang2021CausalSubgroupPSW,Dong2020SBPS}. Related work has also compared estimating propensity scores globally versus within prespecified subsets, highlighting the practical tension between improved local balance and increased variance or numerical instability in small or imbalanced subgroups \cite{Chatelet2023PrespecifiedSubgroupPS}. However, these approaches generally begin with subgroups that are known in advance, such as clinically defined strata or prespecified subgroups. In contrast, the setting motivating our work is one in which clinically relevant heterogeneity may be partially reflected in multivariable pre-treatment covariate patterns but may not be fully captured by a single prespecified subgroup variable. Our contribution is therefore to provide a pragmatic clustering-augmented IPW framework that first learns treatment- and outcome-blind empirical subgroups from baseline covariates, then performs localized propensity-score weighting within these more homogeneous strata, and finally aggregates cluster-specific estimates to obtain an overall precision-weighted summary effect while retaining cluster profiles for diagnostic and clinical interpretation.

A motivating example from oncology illustrates these challenges.
Carboplatin, a platinum-based chemotherapy widely used in breast cancer treatment, can trigger acute hypersensitivity reactions (HSRs), especially after repeated cycles. Clinicians must balance the therapeutic benefit of additional cycles against the risk of severe allergic reactions, but patients who receive more cycles often differ systematically from those who receive fewer cycles in terms of disease severity, prior therapies, and tolerance. These differences confound the relationship between treatment intensity (number of carboplatin cycles) and HSR risk. Standard IPW approaches may produce biased or unstable estimates when such latent heterogeneity is present.

In this study, we propose a Clustering + IPW framework that integrates unsupervised clustering of baseline covariates with localized IPW estimation. Our method first identifies latent patient subgroups based on demographic and clinical characteristics, without using treatment or outcome information, to ensure that clusters represent pre-treatment heterogeneity. IPW is then applied within each cluster, yielding treatment effect estimates that are subsequently combined through inverse-variance weighting. This approach leverages the strengths of both clustering and weighting: it reduces reliance on global parametric assumptions, mitigates extreme weights, and improves interpretability by profiling the characteristics of each cluster.

We evaluate the proposed framework through simulations with and without embedded latent cluster structure and under correctly specified and omitted-covariate propensity-score models. In addition to standard IPW, we consider a natural cluster-informed comparator that includes estimated cluster labels as covariates in a single global propensity-score model. This comparison distinguishes the value of fitting separate treatment-assignment models within clusters from the value of simply incorporating estimated cluster membership into a global model. Across the simulations, the two cluster-informed approaches performed similarly overall, although Clustering + IPW additionally permits treatment-assignment relationships to vary across clusters and produces cluster-specific estimates that may support subgroup interpretation. We then apply the approach to a real-world dataset of 966 breast cancer patients treated with carboplatin at Northwestern Memorial Hospital and Lurie Cancer Center, examining the causal effect of the number of treatment cycles on HSR risk.

The remainder of this manuscript is organized as follows: the Methods section describes the clustering algorithms, weighting techniques, and evaluation criteria used in this study. The Simulation section compares Clustering + IPW with standard IPW, Global PS + Cluster Label, and the naive estimator under various scenarios. The Application section applies the proposed methodology to carboplatin reaction data for breast cancer patients, demonstrating its effectiveness in real-world clinical research. Finally, the Discussion section interprets our findings, acknowledges limitations, and outlines potential directions for future research.

\section{Methods}

In observational studies, estimating the causal effect of a treatment is challenging due to confounding. Propensity score-based methods such as IPW are widely used to adjust for confounding under the assumption that treatment assignment is strongly ignorable given observed covariates. However, when the propensity score model is misspecified, IPW can yield biased and unstable estimates. Besides that, latent subgroups within the population may influence both treatment assignment and outcomes, violating the assumption of homogeneity.

To address these issues, we propose a two-stage framework that combines unsupervised clustering of baseline covariates with IPW estimation. The key idea is to partition the sample into relatively homogeneous clusters that reflect latent subgroup structures, and then estimate propensity scores and treatment effects within each cluster. This Clustering + IPW approach aims to improve covariate balance, enhance robustness to model mis-specification, and account for treatment effect heterogeneity across latent subgroups. The final treatment effect is obtained by aggregating cluster-specific estimates using inverse-variance weighting.

\subsection*{Data Structure and Notation}

Let $i = 1, \ldots, n$ index individuals in the study. For each individual, we observe a vector of baseline covariates $\boldsymbol{X}_i$, a binary treatment indicator $Z_i \in \{0,1\}$, and an outcome $Y_i$. The goal is to estimate the average treatment effect (ATE), defined as $\tau = \mathbb{E}[Y(1) - Y(0)]$, where $Y(1)$ and $Y(0)$ denote the potential outcomes under treatment and control, respectively.

We assume the standard causal inference conditions: (i) consistency, (ii) positivity, and (iii) strong ignorability, i.e., $(Y(1), Y(0)) \perp Z \mid X$.

\subsection*{Step 1: Clustering on Baseline Covariates}

In the first step, we apply an unsupervised learning algorithm to cluster individuals based on their pre-treatment covariate profiles. The goal of this step is to capture latent population structure, such as unmeasured or partially measured subgroups that may affect both treatment assignment and outcomes.

Let $\hat{c}_i \in \{1, \ldots, K\}$ denote the cluster membership for individual $i$, estimated using a clustering method such as hierarchical clustering. Clustering is performed solely on baseline covariates $\boldsymbol{X}_i$, excluding any variables that may be influenced by treatment to avoid conditioning on post-treatment information.

For example, let $\textbf{X}_i = (x_{i1}, x_{i2}, \ldots, x_{ip})^\top$ denote the $p$-dimensional baseline covariate vector for individual $i$, where $\textbf{X}_i$ includes only pre-treatment variables to ensure temporal validity for causal inference. We assume that the study population contains $K$ latent subgroups that differ in both the treatment assignment mechanism and potential outcome distributions. These latent subgroups are not directly observed but can be inferred from observed covariate patterns.

To approximate this latent structure, we apply an unsupervised clustering algorithm to the covariate matrix $\mathbf{X} = (X_1, X_2, \ldots, X_n)^\top$. $\hat{c}_i \in \{1, 2, \ldots, K\}$ denotes the estimated cluster assignment for individual $i$. The clustering algorithm seeks to partition the sample into $K$ clusters such that individuals within the same cluster exhibit high similarity in covariates, while those in different clusters are dissimilar. Take K-means clustering for example, it solves the following optimization problem:
\begin{align}
\min \sum_{k=1}^K \sum_{i= 1}^n \|\textbf{X}_i^{k} - \boldsymbol{\mu}_k\|^2, \nonumber
\end{align}
where $\textbf{X}_i^{k} \in \mathbb{R}^p$ denotes a vector of $p$ baseline covariates for individual $i$ in cluster $k$, $\boldsymbol{\mu}_k \in \mathbb{R}^p$ is the centroid of cluster $k$, and $\|\cdot\|$ denotes the Euclidean norm. The Clustering is performed by minimizing the within-cluster sum of squared distances between individuals and their respective cluster centroids.

Let $\mathcal{I}_k$ be the index set of individuals in cluster $k$, and let $n_k = |\mathcal{I}_k|$ denote the size of cluster $k$. After clustering, the original dataset is partitioned into $K$ non-overlapping subgroups $\{\mathcal{I}_1, \ldots, \mathcal{I}_K\}$ such that $\cup_{k=1}^K \mathcal{I}_k = \{1, 2, \ldots, n\}$ and $\mathcal{I}_k \cap \mathcal{I}_{k'} = \emptyset$ for $k \neq k'$.

Choosing the optimal number of clusters is a critical step in the Clustering + IPW framework because it directly impacts covariate balance, weight stability, and ultimately the estimated treatment effect. We adopt a data-driven approach to select K, using established cluster validity indices such as the average silhouette width to assess the separation and cohesion of clusters. These metrics evaluate how well individuals fit within their assigned clusters compared to other clusters, providing quantitative guidance for determining K.

In addition to statistical criteria, we incorporate practical considerations specific to causal inference. For example, we monitor the size of each cluster to avoid creating subgroups that are too small to support stable propensity score estimation. We also examine the distribution of inverse probability weights after clustering, favoring solutions that yield stable, well-behaved weights with minimal extreme values. In scenarios where multiple K values perform similarly on statistical criteria, we select the solution that balances covariate homogeneity with sufficient within-cluster sample sizes for reliable estimation. This pragmatic strategy is consistent with recommendations in the literature for clustering in causal inference settings \cite{Kassambara2017ClusterAnalysisR}. By integrating both cluster validity indices and downstream diagnostic checks, our method ensures that the chosen number of clusters supports robust and interpretable causal effect estimation.

The outputs from this step, the cluster assignments, directly inform the subsequent stages of the analysis. In Step 2, these cluster labels are used to fit separate propensity score models and compute inverse probability weights within each subgroup. In Step 3, cluster-specific treatment effects are aggregated using inverse-variance weighting, leveraging the stratified design to improve overall estimation accuracy.

\subsection*{Step 2: IPW Estimation Within Clusters}

Within each estimated cluster $\hat{c} \in \{1, \ldots, K\}$, we fit a propensity score model to estimate the probability of treatment conditional on baseline covariates:
\begin{align}
\hat{\pi}_i = {P}(Z_i = 1 \mid X_i), \quad \text{for } i \in \hat{c}. \nonumber
\end{align}

We then compute the standard inverse probability weights:
\begin{align}
w_i = \frac{Z_i}{\hat{\pi}_i} + \frac{1 - Z_i}{1 - \hat{\pi}_i}. \nonumber
\end{align}

Using these weights, we compute the cluster-specific IPW estimator of the treatment effect:
\begin{align}
\hat{\tau}_{\hat{c}} = \frac{\sum_{i \in \hat{c}} w_i Z_i Y_i}{\sum_{i \in \hat{c}} w_i Z_i} - \frac{\sum_{i \in \hat{c}} w_i (1 - Z_i) Y_i}{\sum_{i \in \hat{c}} w_i (1 - Z_i)}. \nonumber
\end{align}


While the Clustering + IPW approach improves robustness to global model misspecification, it introduces additional challenges when cluster sizes are small. Specifically, within-cluster propensity score models may yield extreme or unstable inverse probability weights if treatment assignment is near-deterministic in a particular subgroup or if overlap is limited. Such instability threatens the positivity assumption and can inflate variance, undermining causal effect estimation.

To address this, we implemented adaptive truncation of IPW weights within each cluster, following the strategy proposed by Gruber et al \cite{Gruber2022}. Specifically, to curb the influence of extreme weights, we adopted the \emph{sample–size–adaptive} truncation rule. For each cluster \(c\) with \(n_c\) subjects, the weights were truncated to lie within 
\[
  \text{LB}_c \;=\; \frac{5}{\sqrt{n_c}\,\log n_c},
  \qquad
  \text{UB}_c \;=\; \frac{\sqrt{n_c}\,\log n_c}{5},
\]

where \(\text{LB}_c\) and \(\text{UB}_c\) denote the lower and upper bounds, respectively. Because these limits are determined solely by the cluster’s sample size, the procedure is fully data-adaptive—no fixed quantile or user-chosen tuning constant is required—yet it ensures finite, well-behaved weights that respect the cluster-specific distribution of the generalized propensity scores.

We also conducted diagnostic checks of the weight distributions within each cluster, calculating descriptive statistics such as the mean, standard deviation, and coefficient of variation of the weights. In cases where clusters exhibited near-deterministic treatment assignment (e.g., $>$95\% of patients receiving the same treatment), we flagged these as potential positivity violations and considered combining adjacent clusters to stabilize estimation. This pragmatic approach preserves the benefits of local modeling while mitigating the risk of extreme weights in small or unbalanced subgroups.

\subsection*{Step 3: Aggregation Across Clusters}

After obtaining treatment effect estimates within each cluster, we aggregate the cluster-specific effects to obtain an overall precision-weighted summary effect. We adopt an inverse-variance weighted average:
\begin{align}
\hat{\tau}_{\text{Cluster+IPW}} = \frac{\sum_{\hat{c}} w_{\hat{c}} \hat{\tau}_{\hat{c}}}{\sum_{\hat{c}} w_{\hat{c}}}, \quad \text{where } w_{\hat{c}} = \frac{1}{\widehat{\text{Var}}(\hat{\tau}_{\hat{c}})}. \nonumber
\end{align}

This approach gives more weight to clusters with higher precision and accounts for heterogeneity in sample size and treatment effect variability. If the number of clusters is large or clusters are unbalanced, alternative aggregation strategies such as random-effects meta-analysis may be considered.

For benchmarking, we also implement standard IPW across the full sample. The global propensity score model is estimated without accounting for latent cluster structure. The overall treatment effect is then estimated using:
\begin{align}
\hat{\tau}_{\text{IPW}} = \frac{\sum_i w_i Z_i Y_i}{\sum_i w_i Z_i} - \frac{\sum_i w_i (1 - Z_i) Y_i}{\sum_i w_i (1 - Z_i)}. \nonumber
\end{align}

As an additional cluster-informed comparator, we fit a single global propensity-score model that includes the estimated cluster label as a categorical covariate:
\begin{align}
\operatorname{logit}\{\Pr(Z_i=1\mid \boldsymbol{X}_i,\hat c_i)\}
=
\alpha_0+\boldsymbol{\alpha}^{\top}\boldsymbol{X}_i+
\sum_{k=2}^{K}\delta_k I(\hat c_i=k).
\nonumber
\end{align}
The same estimated cluster assignments and selected value of \(K\) used for Clustering + IPW were used for this comparator. IPW weights and the ATE were then calculated using the same weighted mean contrast as standard IPW. This global model permits cluster-specific intercept shifts while retaining common covariate coefficients across clusters. In contrast, fitting separate propensity-score models within clusters permits both the intercepts and covariate coefficients to vary across clusters.

Standard IPW and Global PS + Cluster Label serve as simulation comparators for the proposed Clustering + IPW estimator. We compare the estimators based on bias, empirical variability, mean squared error, and confidence-interval coverage. In the carboplatin application, standard IPW serves as the primary benchmark.

\section{Simulation}

To evaluate the performance of the proposed Clustering + IPW approach, we conducted a simulation study under two distinct scenarios and compared four estimators: the naive estimator, standard IPW, Clustering + IPW, and Global PS + Cluster Label. In Scenario A, treatment assignment and outcomes were generated from baseline covariates without embedded latent cluster structure. Clustering was applied post hoc to evaluate the consequences of partitioning the covariate space when no true subgroups were present. In Scenario B, latent clusters influenced the covariate distributions, treatment-assignment intercepts, and outcome intercepts, creating cluster structure that was partially recoverable from observed covariate patterns. Across both scenarios, we evaluated bias, empirical variability, mean squared error (MSE), and confidence-interval coverage under correctly specified and omitted-covariate propensity-score models.

In each simulation iteration, we selected the number of clusters \(K\) data‐adaptively by maximizing the average silhouette width. Specifically, after generating the covariate, treatment, and outcome data for that replicate, we applied agglomerative hierarchical clustering (Ward’s linkage on Gower dissimilarity) for every \(K \in \{2,3,4,5\}\). The value of \(K\) that achieved the largest silhouette width was retained for the Clustering + IPW analysis. Unlike the carboplatin application, we did not impose an additional weight‐stability screen, as inspecting weight distributions across thousands of replicates would be computationally prohibitive and difficult to automate.
Figure \ref{fig:k_dist} shows how often the silhouette criterion recovered the true \(K = 3\) in the latent‐cluster scenario, with accuracy improving from roughly 60\% at \(n = 100\) to over 80\% for \(n \ge 300\). To assess the impact of mis‐specifying \(K\), we conducted a sensitivity analysis in which the clustering step was fixed at \(K = 2\) or \(K = 4\). Results in Figure~\ref{fig:mse_misK} indicate that both cluster-informed estimators retained substantial improvements over misspecified standard IPW when \(K\) was fixed at 2 or 4, although neither cluster-informed approach uniformly dominated the other.

\subsection*{Design}

The first scenario considers a population with no latent clusters, while the second incorporates latent cluster structure that influences both treatment assignment and outcomes. Across both settings, we evaluated the relative bias, empirical variability, MSE, and confidence-interval coverage of the four estimators in estimating the average treatment effect (ATE). For each scenario, we considered sample sizes \( n \in \{100, 200, 300, 400, 500\} \) and generated \( R = 1000 \) replicates.

Our estimand is the ATE, defined as \( \tau = \mathbb{E}[Y(1)-Y(0)] \), with the true treatment effect fixed at \(\beta_1=1.5\). We compared four estimators: (i) the naive difference-in-means estimator; (ii) standard IPW, in which one propensity-score model is fit to the full sample; (iii) Clustering + IPW, which fits separate propensity-score models within estimated clusters and combines the cluster-specific estimates using inverse-variance weighting; and (iv) Global PS + Cluster Label, which includes the estimated cluster membership as a categorical covariate in one global propensity-score model. For the three IPW-based estimators, we compute weights
\[
w_i \;=\; \frac{Z_i}{\hat{\pi}(X_i)} \;+\; \frac{1 - Z_i}{1 - \hat{\pi}(X_i)},
\]

For the correctly specified PS analyses, \(\hat{\pi}(X_i)\) was estimated using a logistic regression including all three covariates used in the treatment assignment mechanism. Specifically, in Scenario A (No Latent Clusters), the fitted PS model was
\[
\text{logit}\{\hat{\pi}(X_i)\} = \alpha_0 + \alpha_1 x_{i1} + \alpha_2 x_{i2} + \alpha_3 x_{i3}.
\]
In Scenario B (Latent Clusters Embedded), the same covariates were included; for the standard IPW estimator, this model was fit once in the full sample, while for Clustering + IPW it was refit separately within each estimated cluster, allowing the coefficients to vary across clusters through the within-cluster fitting procedure. To induce PS model misspecification, we fit a reduced logistic regression that included only \(x_{i1}\), intentionally omitting \(x_{i2}\) and \(x_{i3}\), even though both omitted covariates contributed to the true treatment assignment mechanisms in Scenarios A and B:
\[
\text{logit}\{\hat{\pi}_{\text{mis}}(X_i)\} = \alpha_0 + \alpha_1 x_{i1}.
\]
This reduced PS model was used for both the misspecified standard IPW estimator and the misspecified Clustering + IPW estimator, with the latter refit separately within each estimated cluster. Thus, PS misspecification in the simulations represents omitted-covariate misspecification rather than misspecification of the link function. Performance was assessed using the empirical bias, variance, and MSE of the ATE estimates across replicates.

For Global PS + Cluster Label, the correctly specified model included \(x_{i1}\), \(x_{i2}\), \(x_{i3}\), and the estimated cluster indicators. Its misspecified version included only \(x_{i1}\) and the estimated cluster indicators. Thus, the same omitted-covariate misspecification was applied to all propensity-score estimators.

\subsection*{Data Generating Mechanisms}

In both data generating mechanisms, each subject \( i = 1, \ldots, n \) has three baseline covariates \( X_i = (x_{i1}, x_{i2}, x_{i3}) \) that drive both treatment assignment and outcome. Potential outcomes are generated according to
\[
Y_i(z) = \beta_0 + \beta_1 z + \beta_2 x_{i1} + \beta_3 x_{i2} + \beta_4 x_{i3} + \varepsilon_i,
\]
where \( \varepsilon_i \sim \mathcal{N}(0, 1) \), with parameters \( (\beta_0, \beta_1, \beta_2, \beta_3, \beta_4) = (2.0, 1.5, 1.0, -1.0, 0.5) \). The observed outcome is \( Y_i = Y_i(Z_i) \).

\textbf{Scenario A (No Latent Clusters).}
Covariates are drawn independently from standard normal distributions \( x_{ij} \sim \mathcal{N}(0, 1) \). Treatment assignment follows a logistic model
\[
\text{logit}\{\Pr(Z_i = 1 \mid X_i)\} = 0.20 + 0.40 x_{i1} - 0.30 x_{i2} + 0.50 x_{i3}.
\]
Although no true subgroups exist, we apply hierarchical clustering to \( (x_{i1},x_{i2},x_{i3}) \) to evaluate whether partitioning the observed covariate space improves or degrades propensity-score estimation when no explicit latent structure is present.

\textbf{Scenario B (Latent Clusters Embedded).}
To represent covariate-recoverable population heterogeneity, each individual is first assigned to one of \( C = 3 \) latent clusters according to \( (p_1, p_2, p_3) = (0.30, 0.40, 0.30) \). Conditional on cluster membership \( c_i \), covariates are generated from cluster-specific distributions:
\[
x_{i1} \sim \mathcal{N}(\mu_{1c_i}, 1), \quad
x_{i2} \sim \mathcal{N}(\mu_{2c_i}, 1), \quad
x_{i3} \sim \mathcal{N}(\mu_{3c_i}, 1),
\]
with cluster means \( \boldsymbol{\mu}_1 = (-1, 0, 1), \boldsymbol{\mu}_2 = (0, 1, -1), \boldsymbol{\mu}_3 = (1, -1, 0) \). Treatment assignment is modeled as
\[
\text{logit}\{\Pr(Z_i = 1 \mid c_i, X_i)\} = \gamma_{0c_i} + 0.40 x_{i1} - 0.30 x_{i2} + 0.50 x_{i3},
\]
where \( \gamma_{0c_i} \in \{0.10, -0.10, 0.20\} \) varies by latent cluster, and outcomes follow the same form as above but with cluster-specific intercepts \( \beta_{0c_i} \in \{1.5, 2.0, 1.0\} \). This scenario evaluates the estimators when subgroup structure affecting treatment assignment and outcome levels is reflected in the observed covariate distributions.

The following section reports findings from 1,000 simulation replicates under the two primary scenarios. A focused comparison of the two cluster-informed estimators is presented in Table~\ref{tab:cluster_comparator}, complete numerical results are provided in Supplementary Tables S1 and S2, and MSE results across all sample sizes are visualized in Figures~\ref{fig:mse_wCluster} and \ref{fig:mse_woCluster}.

\subsection*{Simulation Results}

\FloatBarrier
\begin{table*}[t]
\centering
\caption{Comparison of Clustering + IPW and Global PS + Cluster Label at the smallest and largest sample sizes.}
\label{tab:cluster_comparator}
\small
\setlength{\tabcolsep}{5pt}
\begin{tabular}{llrccc|ccc}
\toprule
& & & \multicolumn{3}{c|}{Clustering + IPW} &
\multicolumn{3}{c}{Global PS + Cluster Label} \\
Scenario & PS specification & \(n\) & Bias (\%) & MSE & Coverage &
Bias (\%) & MSE & Coverage \\
\midrule
Latent clusters
 & Correct      & 100 & 9.0  & 0.134 & 0.883 & 2.5  & 0.140 & 0.947 \\
 & Correct      & 500 & 0.2  & 0.016 & 0.955 & -2.0 & 0.022 & 0.957 \\
 & Misspecified & 100 & 29.0 & 0.323 & 0.657 & 27.8 & 0.285 & 0.727 \\
 & Misspecified & 500 & 27.9 & 0.203 & 0.168 & 27.4 & 0.198 & 0.200 \\
\midrule
No latent clusters
 & Correct      & 100 & 7.3  & 0.112 & 0.879 & 1.0  & 0.079 & 0.964 \\
 & Correct      & 500 & 1.4  & 0.011 & 0.945 & 0.0  & 0.011 & 0.958 \\
 & Misspecified & 100 & 20.4 & 0.212 & 0.710 & 18.8 & 0.178 & 0.820 \\
 & Misspecified & 500 & 22.3 & 0.135 & 0.268 & 22.8 & 0.140 & 0.276 \\
\bottomrule
\end{tabular}

\vspace{0.5ex}
{\footnotesize Bias is reported as a percentage of the true treatment effect, \(\tau=1.5\). Complete results across all sample sizes and estimators are reported in Supplementary Tables S1 and S2.}
\end{table*}

\begin{figure}[t]
    \centering
    \includegraphics[width=0.9\columnwidth]{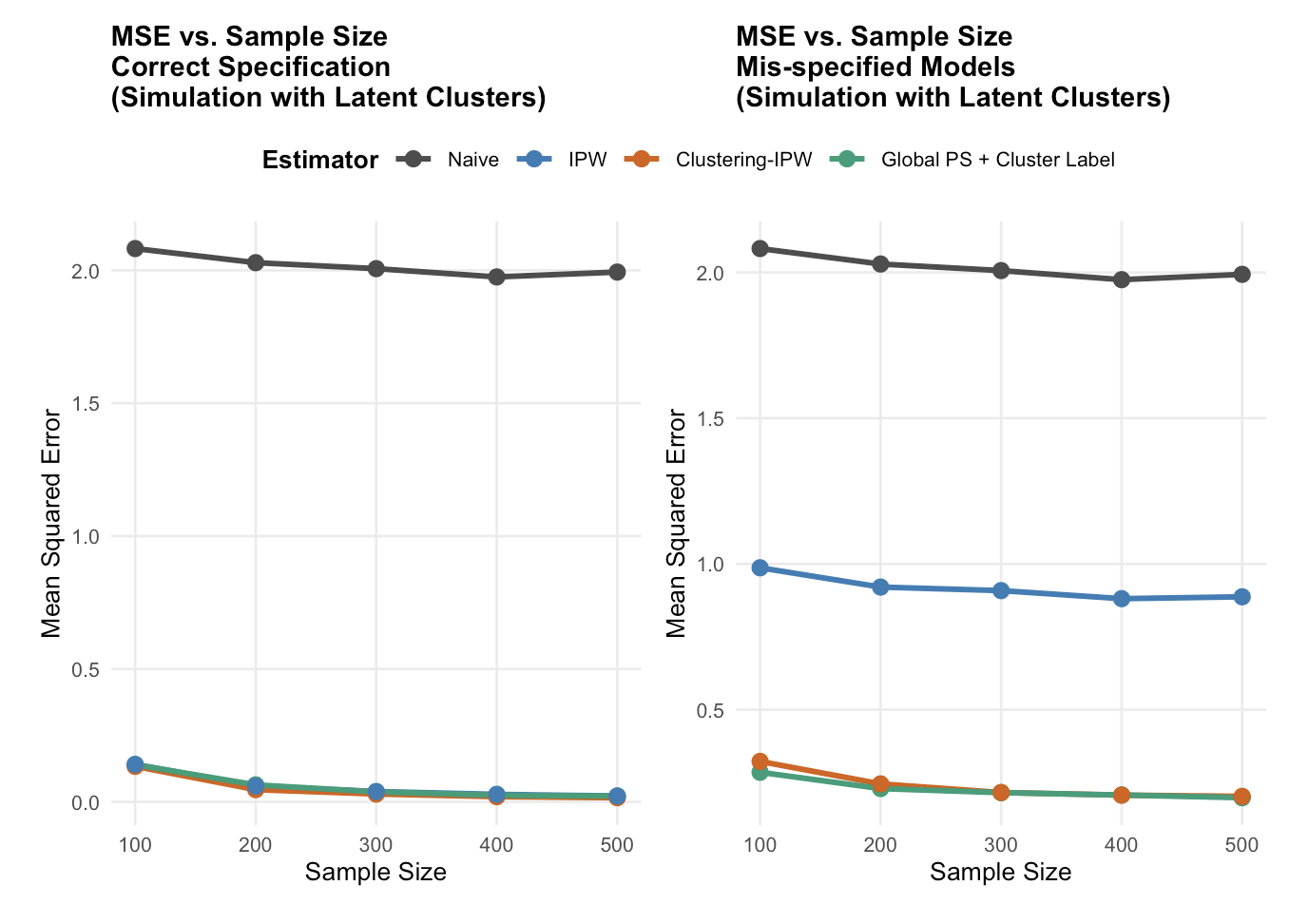}
    \caption{Mean squared error across sample sizes in the latent-cluster scenario for the naive estimator, standard IPW, Clustering + IPW, and Global PS + Cluster Label. The left panel shows results under correct propensity-score specification; the right panel shows results under omitted-covariate misspecification.}
    \label{fig:mse_wCluster}
\end{figure}

\begin{figure}[t]
    \centering
    \includegraphics[width=0.9\columnwidth]{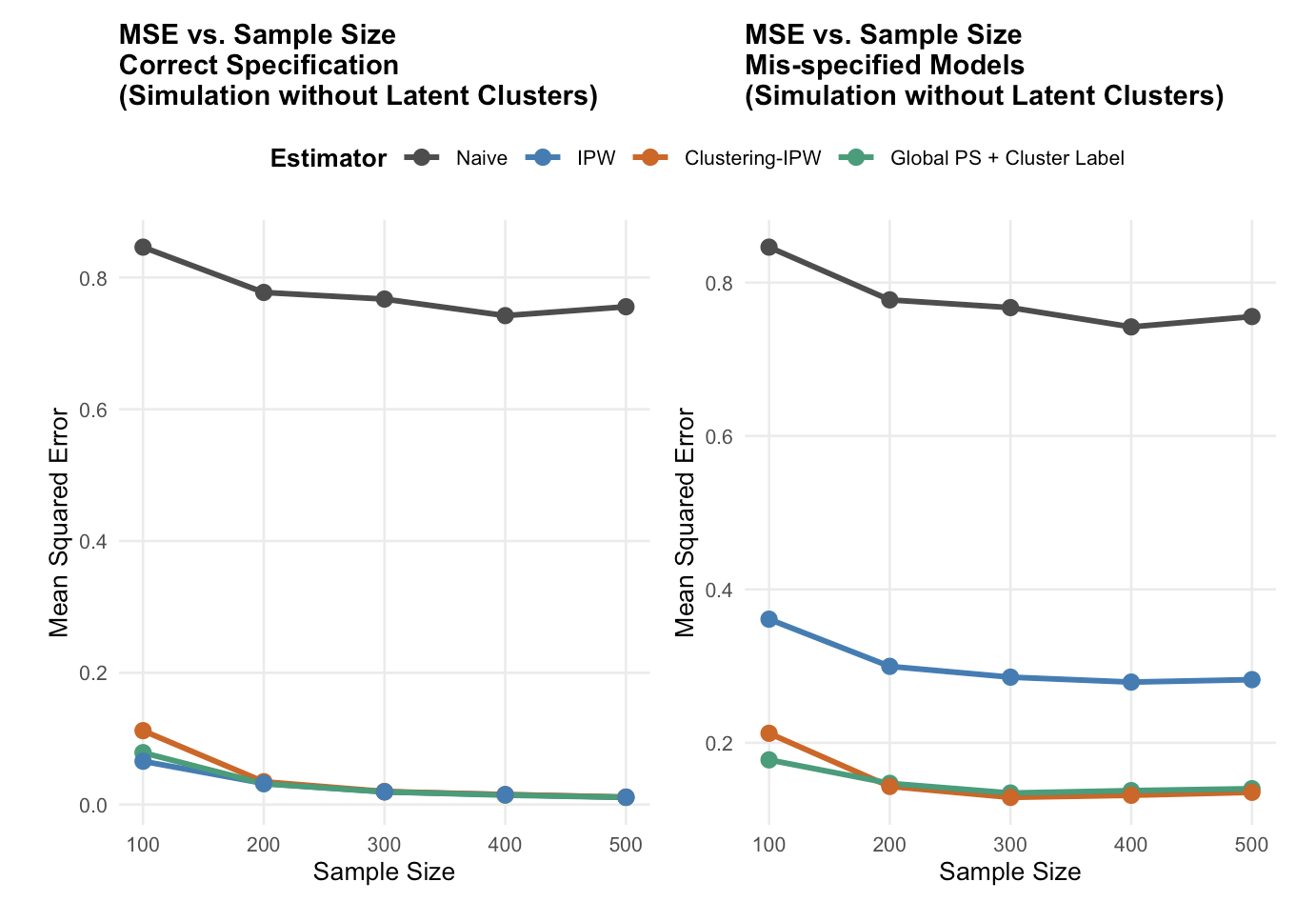}
    \caption{Mean squared error across sample sizes in the scenario without embedded latent clusters for the naive estimator, standard IPW, Clustering + IPW, and Global PS + Cluster Label. The left panel shows results under correct propensity-score specification; the right panel shows results under omitted-covariate misspecification.}
    \label{fig:mse_woCluster}
\end{figure}

\FloatBarrier
Table~\ref{tab:cluster_comparator} provides a focused comparison of Clustering + IPW and Global PS + Cluster Label, while Supplementary Tables S1 and S2 report complete results for all four estimators. Figures~\ref{fig:mse_wCluster} and \ref{fig:mse_woCluster} display MSE across all evaluated sample sizes. The two primary scenarios used 1,000 replicates for each sample size \(n\in\{100,200,300,400,500\}\).

In the latent-cluster scenario under correct propensity-score specification, Clustering + IPW achieved lower MSE than Global PS + Cluster Label at every evaluated sample size. Global PS + Cluster Label had lower absolute bias at \(n=100\), \(200\), and \(300\), whereas Clustering + IPW had lower absolute bias at \(n=400\) and \(500\). Global PS + Cluster Label had better coverage across all sample sizes, although the difference was minimal at \(n=500\) (0.957 versus 0.955). These findings demonstrate a tradeoff between MSE, bias, and coverage rather than uniform superiority of either cluster-informed estimator.

Under propensity-score misspecification in the latent-cluster scenario, standard IPW exhibited approximately 62\% bias and severe undercoverage. Clustering + IPW and Global PS + Cluster Label reduced the bias to approximately 27--29\% and substantially reduced MSE. At \(n=100\), their MSEs were 0.323 and 0.285, respectively, compared with 0.987 for standard IPW. At \(n=500\), the corresponding MSEs were 0.203, 0.198, and 0.887. The two cluster-informed estimators therefore performed similarly overall, but neither eliminated the bias or undercoverage caused by omitted-covariate misspecification.

In the scenario without embedded latent clusters, Global PS + Cluster Label was generally preferable to Clustering + IPW under correct specification, particularly at smaller sample sizes. At \(n=100\), the MSEs were 0.079 for Global PS + Cluster Label, 0.112 for Clustering + IPW, and 0.066 for standard IPW. By \(n=500\), all three adjusted estimators had MSEs of approximately 0.011. Thus, fitting separate propensity-score models within empirically generated clusters did not provide an efficiency advantage when the global propensity-score model was correctly specified and no latent cluster structure was present.

Under misspecification without latent clusters, both cluster-informed estimators substantially improved over misspecified standard IPW. At \(n=100\), MSE was 0.212 for Clustering + IPW and 0.178 for Global PS + Cluster Label, compared with 0.361 for standard IPW. At \(n=500\), the corresponding values were 0.135, 0.140, and 0.282. Differences between the two cluster-informed approaches were modest and changed direction across sample sizes, providing no evidence of uniform superiority.

Confidence-interval coverage was generally near nominal under correct propensity-score specification for standard IPW and Global PS + Cluster Label, whereas Clustering + IPW showed some undercoverage at smaller sample sizes. Under misspecification, both cluster-informed estimators improved coverage relative to standard IPW, particularly at smaller sample sizes, but coverage declined as sample size increased. For example, in the latent-cluster scenario at \(n=500\), coverage was 0.168 for Clustering + IPW, 0.200 for Global PS + Cluster Label, and 0 for standard IPW. These findings indicate that cluster-informed adjustment reduced bias-driven undercoverage in the studied settings but did not resolve propensity-score misspecification.

\begin{figure}[t]
    \centering
    \includegraphics[width=0.85\columnwidth]{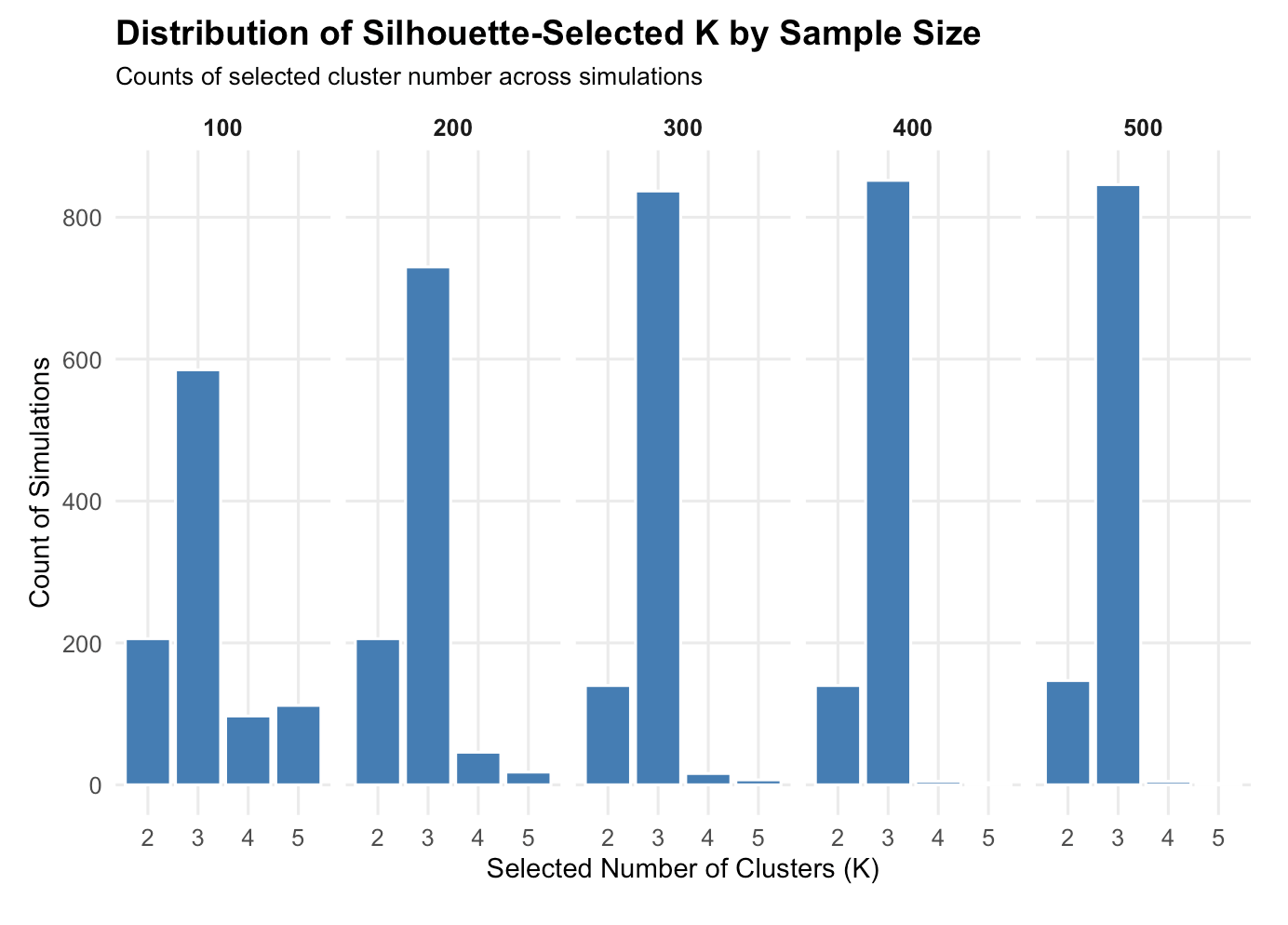}
    \caption{Distribution of Silhouette‐Selected \(K\) Across 1{,}000 Latent‐Cluster Simulations.  
             Each panel shows, for a given sample size, how many replicates
             chose \(K \in \{2,3,4,5\}\) as the silhouette‐width “optimal’’
             number of clusters.  The true latent structure contained  
             three clusters.}
    \label{fig:k_dist}
\end{figure}

Figure~\ref{fig:k_dist} summarizes how often the silhouette criterion identified the correct number of clusters (\(K=3\)) in Scenario B of the simulation study (latent‐cluster embedded scenario).  When the sample size was small (\(n=100\)), the silhouette rule selected the true value in \(\approx 60\%\) of replicates, most of the remaining runs collapsing to \(K=2\).  As sample size increased, the proportion of correct selections rose steadily: about \(75\%\) at \(n=200\) and exceeding \(80\%\) for \(n\ge 300\).  Selections of \(K\ge 4\) were rare across all sample sizes.  This pattern illustrates the expected improvement in cluster‐number recovery with larger samples—the additional data sharpen the silhouette contrast, reducing the tendency to merge or split clusters when observations are sparse.

\subsubsection*{Sensitivity Analysis: Mis-Specification of the Number of Clusters}

\begin{figure}[t]
    \centering
    \includegraphics[width=0.85\columnwidth]{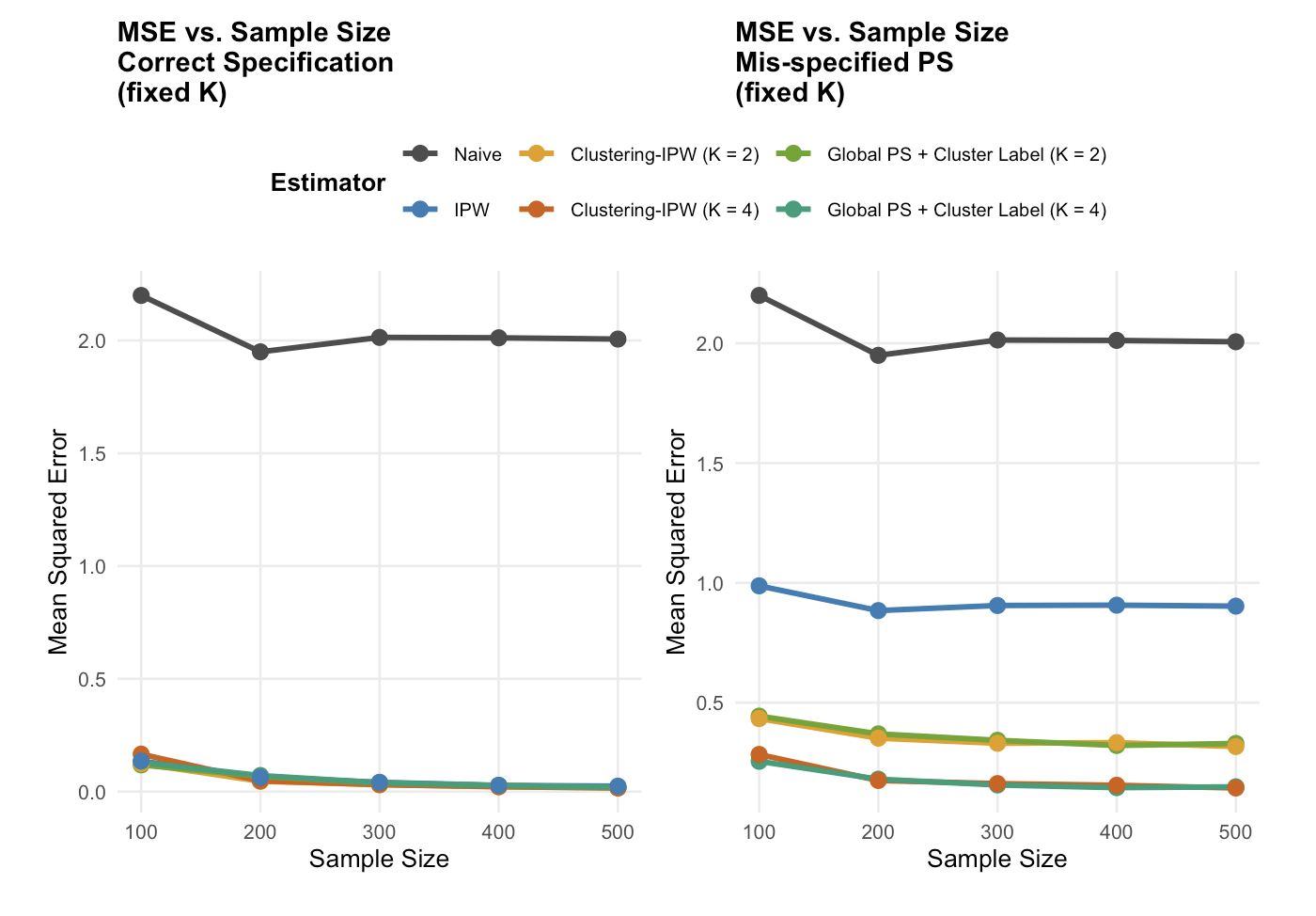}
    \caption{Mean squared error in the fixed-\(K\) sensitivity analysis based on 500 replicates per sample size. Clustering + IPW and Global PS + Cluster Label were evaluated with \(K=2\) and \(K=4\), while the true latent structure contained \(K=3\). The left panel shows correct propensity-score specification; the right panel shows omitted-covariate misspecification.}
    \label{fig:mse_misK}
\end{figure}

Although the optimal number of clusters was chosen by silhouette width in the main simulation, we conducted a sensitivity analysis in which the clustering step was deliberately misspecified by fixing \(K=2\) and \(K=4\). Figure~\ref{fig:mse_misK} shows the resulting MSE curves.

\begin{itemize}
  \item \textbf{Correct PS (left panel).}  
        Under correct propensity-score specification, Global PS + Cluster Label generally had lower bias and better coverage than Clustering + IPW, particularly at \(n=100\). At moderate-to-large sample sizes, Clustering + IPW generally achieved similar or lower MSE.
  \item \textbf{misspecified PS (right panel).}  
        Under misspecification, the two cluster-informed estimators again performed similarly within each value of \(K\). The \(K=4\) implementations generally had lower MSE than the corresponding \(K=2\) implementations, while both substantially improved over misspecified standard IPW. No estimator uniformly dominated across all combinations of \(K\), sample size, and propensity-score specification.
\end{itemize}

These findings suggest that moderate over- or under-partitioning did not eliminate the performance gains of cluster-informed adjustment relative to misspecified standard IPW. However, misspecifying \(K\) is not equivalent to failing to recover treatment-assignment-relevant cluster structure, and neither cluster-informed estimator uniformly dominated the other.

\section*{Application}

Carboplatin is a widely used platinum-based chemotherapeutic agent for the treatment of breast and gynecologic malignancies. While generally well tolerated, repeated administration of carboplatin is associated with an increased risk of acute hypersensitivity reactions (HSRs), which can range from mild allergic symptoms to life-threatening anaphylaxis. These reactions are believed to be immunologically mediated and are more common after multiple cycles of carboplatin exposure. Understanding the causal effect of treatment intensity, particularly the number of carboplatin cycles administered, on the risk of hypersensitivity is essential for guiding dosing strategies, premedication protocols, and clinical decision-making in oncology practice.

However, estimating this effect from observational data poses several challenges. Patients who receive more cycles of carboplatin often differ systematically from those who receive fewer cycles in terms of prior treatment history (e.g., exposure to other chemotherapy agents), treatment response, and clinical tolerance. These differences may confound the observed association between carboplatin exposure and HSR risk. Moreover, standard methods such as inverse probability weighting (IPW) may be vulnerable to model misspecification in the presence of latent patient heterogeneity. We address this challenge by applying the proposed Clustering + IPW framework, which stratifies the population into latent subgroups based on baseline characteristics and performs localized propensity score adjustment within each subgroup.

The analytic dataset was derived from a retrospective cohort of breast cancer patients treated at Northwestern Memorial Hospital and Lurie Cancer Center. The cohort includes a total of 966 individuals who received at least one dose of carboplatin. Among these, 29 patients (3.0\%) experienced a documented hypersensitivity reaction during their treatment course, as adjudicated by clinical allergists based on symptom profiles, timing, and supportive diagnostic data.

Baseline variables as shown in table \ref{tab:baseline} included demographic characteristics (e.g., age, race, body surface area) and clinical history (e.g., prior exposure to radiation therapy), along with other baseline features used for clustering. Treatment intensity was represented solely by the total number of carboplatin cycles administered.

\FloatBarrier
\begin{table*}[t]
  \centering
  \caption{Baseline characteristics of the study cohort (N = 966).}
  \label{tab:baseline}
  \small
  \setlength{\tabcolsep}{6pt}
  \renewcommand{\arraystretch}{0.92}

  \begin{tabular*}{\textwidth}{@{\extracolsep{\fill}} l r l r @{}}
    \toprule
    \textbf{Characteristic} & \textbf{Value} &
    \textbf{Characteristic} & \textbf{Value} \\
    \midrule

    \multicolumn{2}{l}{\textbf{Status (n \%)}} & \multicolumn{2}{l}{\textbf{Radiation (n \%)}} \\
    \quad Non‐Reactor & 937 (97.0) & \quad No  & 886 (91.7) \\
    \quad Reactor     &  29 (3.0)  & \quad Yes &  80 (8.3)  \\
    \addlinespace[0.5ex]

    \multicolumn{2}{l}{\textbf{Age (mean (SD))}} &
    \multicolumn{2}{l}{\textbf{Treatment intensity}} \\
    \quad & 52.3 (11.9) & \quad Cycles (mean (SD)) & 5.1 (2.7) \\
    \addlinespace[0.5ex]

    \multicolumn{2}{l}{\textbf{BSA (m$^2$) (mean (SD))}} &
    \multicolumn{2}{l}{\textbf{Chemo agents (n \%)}} \\
    \quad & 1.8 (0.2)   & \quad Taxol         & 792 (82.0) \\
                                &               & \quad Pembrolizumab & 184 (19.0) \\
                                &               & \quad Pemetrexed    &   2 (0.2)  \\
                                &               & \quad Trastuzumab   & 588 (60.9) \\
                                &               & \quad Pertuzumab    & 500 (51.8) \\
                                &               & \quad Gemcitabine   & 123 (12.7) \\
    \addlinespace[0.5ex]

    \multicolumn{2}{l}{\textbf{Gender (n \%)}} &
    \multicolumn{2}{l}{\textbf{Stage (n \%)}} \\
    \quad Female (F) & 964 (99.8) & \quad I        & 114 (30.4) \\
    \quad Male (M)   &   2 (0.2)  & \quad II       & 136 (36.3) \\
                     &            & \quad III      &  70 (18.7) \\
                     &            & \quad IV       &  55 (14.7) \\
                     &            & \quad Missing  & 591 \\
    \addlinespace[0.5ex]

    \multicolumn{2}{l}{\textbf{Race (n \%)}} & \multicolumn{2}{l}{} \\
    \quad White & 678 (70.2) & & \\
    \quad Black & 122 (12.6) & & \\
    \quad Asian &  59 (6.1)  & & \\
    \quad Other & 107 (11.1) & & \\
    \bottomrule
  \end{tabular*}

  \vspace{0.3ex}
  {\footnotesize Continuous variables are mean (SD); categorical variables are n (\%).}
\end{table*}
\FloatBarrier

For the purposes of this analysis, the primary exposure was the number of carboplatin cycles, modeled as a continuous predictor. The outcome was a binary indicator of whether the patient experienced an HSR. Clustering was performed using baseline covariates only, excluding both exposure and outcome. The Clustering + IPW analysis was used to estimate the per-cycle dose--response contrast under measured-confounding, positivity, consistency, and correct-model-specification assumptions, with standard IPW serving as the benchmark.

Because cumulative carboplatin exposure is naturally measured on a continuous scale, we explored both dichotomized and continuous representations. Dichotomization substantially reduced information and resulted in sparse within-cluster exposure groups, yielding unstable weighted estimates. We therefore retained the continuous exposure formulation using generalized propensity scores, which better reflects the underlying treatment process while avoiding information loss associated with arbitrary categorization.

Using generalized propensity score (GPS)\cite{Austin2018GPS}, we first estimated the dose–response effect under the traditional IPW framework. The GPS model included baseline covariates selected for their clinical relevance and potential to confound the relationship between treatment intensity and hypersensitivity reaction (HSR) risk. Specifically, the model incorporated demographic variables (age, race categorized as White, Black, Asian, and Other, and body surface area (BSA)), and clinical history (prior exposure to radiation therapy). The treatment variable of interest was the number of carboplatin cycles, and a stabilized weight was computed using the ratio of the marginal density to the conditional GPS\cite{Austin2018GPS}. As described by Gruber et al.\cite{Gruber2022}, we applied a data‐adaptive truncation to the stabilized weights, setting lower and upper bounds that depend on the sample size. This truncation prevents excessively small or large weights, thereby controlling extreme values and enhancing the stability of the IPW estimates. 

As summarized in Table~\ref{tab:weight_trunc_summary_k3}, the untruncated stabilized weights under the \(K=3\) clustering were already well behaved, with means near 1 (0.998--1.007), moderate dispersion (SD 0.126--0.286), and maxima between 1.77 and 4.95. After truncation, the distributions were essentially unchanged. The weights in Clusters 1 and 2 were unaffected, and in Cluster 3 only the minimum increased from 0.018 to 0.033; its mean, SD, and maximum remained unchanged to three decimal places. These results indicate stable weighting across clusters with little influence from truncation.

The IPW-weighted logistic regression model revealed a statistically significant association: each additional carboplatin cycle was associated with an 18.0\% increase in the odds of hypersensitivity (OR \(=1.180\), 95\% CI: 1.097--1.282; \(p<0.001\)).

\begin{table*}[t]
\centering
\caption{Summary of pre- and post-truncation stabilized weights by cluster ($K=3$).}
\label{tab:weight_trunc_summary_k3}
\small
\begin{tabular}{lcccccccc}
\toprule
 & \multicolumn{4}{c}{Pre-truncation} & \multicolumn{4}{c}{Post-truncation} \\
\cmidrule(lr){2-5}\cmidrule(lr){6-9}
Cluster & Mean & SD & Min & Max & Mean & SD & Min & Max \\
\midrule
1 & 0.998 & 0.212 & 0.230 & 2.085 & 0.998 & 0.212 & 0.230 & 2.085 \\
2 & 1.002 & 0.126 & 0.137 & 1.773 & 1.002 & 0.126 & 0.137 & 1.773 \\
3 & 1.007 & 0.286 & 0.018 & 4.949 & 1.007 & 0.286 & 0.033 & 4.949 \\
\bottomrule
\end{tabular}
\end{table*}

\FloatBarrier

Next, we applied the Clustering + IPW approach to address potential latent heterogeneity in treatment assignment and outcome relationships. Hierarchical clustering with Gower dissimilarity, which is well suited to mixed data \cite{Coombes2021}, was conducted using age, gender, race, ethnicity, BSA, cancer stage, prior radiation therapy, and indicators for Taxol, pembrolizumab, pemetrexed, trastuzumab, pertuzumab, and gemcitabine, explicitly excluding cumulative carboplatin exposure and HSR status. Because silhouette widths were comparably high for $K=2,3,4$ and the best separation was achieved at $K=3$, we selected $K=3$ as the final cluster count. Within each cluster, we refit the GPS model using the same baseline covariates as the standard IPW model (age, race, BSA, radiation). Stabilized weights were computed within cluster as the ratio of the marginal to conditional GPS, with data-adaptive truncation to control extremes \cite{Gruber2022}. Cluster-specific IPW-weighted treatment effects were then estimated via logistic regression with HSR as the outcome and number of cycles as the exposure, and the cluster-level log-odds ratios were combined using inverse-variance weighting \cite{higgins2024cochrane,MarinMartinez2010}. The pooled estimate indicated a significant association between cycles and hypersensitivity (IVW OR \(=1.220\), 95\% CI: 1.069--1.391).

\begin{figure}[t]
  \centering
  \includegraphics[width=\columnwidth]{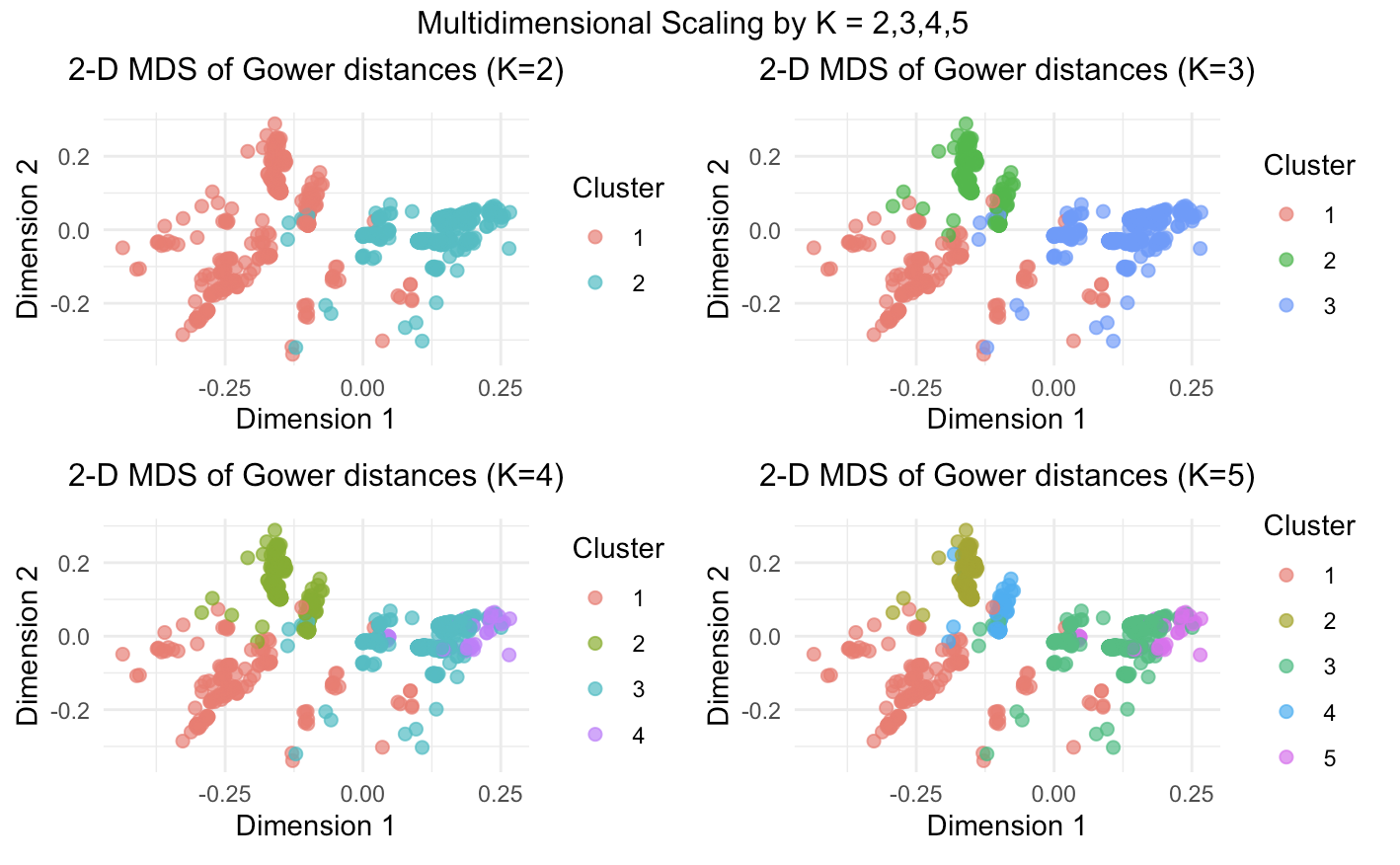}
  \caption{Multidimensional scaling (MDS) projections of Gower distances based on baseline covariates for \(K=2,3,4,5\).}
  \label{fig:mds-all}
\end{figure}

Figure~\ref{fig:mds-all} displays the MDS projections for \(K=2,3,4,5\), illustrating the empirical separation of patient profiles under alternative cluster solutions. The \(K=3\) solution had the highest average silhouette width while retaining adequate cluster sizes and stable weights. These diagnostics support \(K=3\) as a practical partition of the observed covariate space, but they do not by themselves establish clinically distinct treatment effects. Table~\ref{tab:cluster-characteristics-K3} therefore provides descriptive cluster profiles to support interpretation of the empirical subgroups.

\FloatBarrier

Table~\ref{tab:cluster-specific-K3} presents cluster-specific odds-ratio estimates for the association between number of carboplatin cycles and hypersensitivity. Clusters 1 and 2 had estimates above one (Cluster 1: OR \(=1.273\), \(p=0.002\); Cluster 2: OR \(=3.660\), \(p<0.001\)), whereas Cluster 3 had an estimate below one (OR \(=0.690\), \(p=0.020\)). Because only 29 HSR events were observed in the full cohort, these cluster-specific results should be interpreted as exploratory evidence of variation rather than definitive evidence of treatment-effect heterogeneity.

Table~\ref{tab:ivw-ipw-comparison} compares the IVW-pooled estimate across clusters with the standard IPW estimate. The IVW-pooled OR was 1.2199 (95\% CI: 1.0694--1.3915), while the standard IPW OR was 1.1797 (95\% CI: 1.0965--1.2816). Both approaches indicate higher odds of HSR with each additional cycle and produce pooled estimates of similar magnitude. The cluster-specific estimates provide additional descriptive information about empirical patient subgroups but should not be interpreted as confirming heterogeneous treatment effects.

\begin{table*}[t]
\centering
\caption{Cluster-specific dose--response estimates for \(K=3\).}
\label{tab:cluster-specific-K3}
\begin{tabular}{@{}rrrrrrr@{}}
\toprule
Cluster & N & OR & SE (OR) & 95\% CI Lower & 95\% CI Upper & p-value \\
\midrule
1 & 172 & 1.273 & 0.098 & 1.106 & 1.509 & 0.002 \\
2 & 227 & 3.660 & 0.988 & 2.282 & 6.712 & $<0.001$ \\
3 & 567 & 0.690 & 0.110 & 0.508 & 0.969 & 0.020 \\
\bottomrule
\end{tabular}
\end{table*}
\FloatBarrier

\begin{table*}[t]
\centering
\caption{Comparison of IVW-pooled and standard IPW dose--response estimates.}
\label{tab:ivw-ipw-comparison}
\begin{tabular}{l r r r r}
\toprule
Method       & OR     & SE (OR) & 95\% CI Lower & 95\% CI Upper \\
\midrule
IVW-pooled   & 1.2199 & 0.0819 & 1.0694 & 1.3915 \\
Standard IPW & 1.1797 & 0.0454 & 1.0965 & 1.2816 \\
\bottomrule
\end{tabular}
\end{table*}
\FloatBarrier

\begin{table*}[t]
\centering
\caption{Baseline characteristics by cluster ($K=3$).}
\label{tab:cluster-characteristics-K3}
\small
\setlength{\tabcolsep}{4pt}
\renewcommand{\arraystretch}{1.2}
\begin{tabularx}{\textwidth}{
   >{\raggedright\arraybackslash}p{0.22\textwidth}
   >{\raggedright\arraybackslash}p{0.22\textwidth}
   >{\raggedright\arraybackslash}p{0.22\textwidth}
   >{\raggedright\arraybackslash}p{0.22\textwidth}
   >{\centering\arraybackslash}p{0.10\textwidth}
}
\toprule
Variable & Cluster 1 (n=172) & Cluster 2 (n=227) & Cluster 3 (n=567) & p-value \\
\midrule
Age (mean $\pm$ SD)
  & 55.4 $\pm$ 11.8
  & 51.8 $\pm$ 12.9
  & 51.5 $\pm$ 11.4
  & 0.001 \\

Race (\%)
  & \makecell[l]{White: 72.1\% \\ Black: 14.5\% \\ Asian: 4.1\% \\ Other: 9.3\%}
  & \makecell[l]{White: 70.0\% \\ Black: 13.2\% \\ Asian: 4.8\% \\ Other: 11.9\%}
  & \makecell[l]{White: 69.7\% \\ Black: 11.8\% \\ Asian: 7.2\% \\ Other: 11.3\%}
  & 0.603 \\

Prior radiation (\%)
  & 22.7\%
  & 0.4\%
  & 7.1\%
  & $<0.0001$ \\

No.\ of cycles (mean $\pm$ SD)
  & 5.1 $\pm$ 4.9
  & 4.0 $\pm$ 1.5
  & 5.6 $\pm$ 1.8
  & $<0.0001$ \\

Stage distribution (\%)
  & \makecell[l]{I: 0.6\% \quad II: 5.2\% \\ III: 1.2\% \quad IV: 20.9\% \\ Missing: 72.1\%}
  & \makecell[l]{I: 6.2\% \quad II: 19.8\% \\ III: 17.6\% \quad IV: 4.0\% \\ Missing: 52.4\%}
  & \makecell[l]{I: 17.5\% \quad II: 14.5\% \\ III: 4.9\% \quad IV: 1.8\% \\ Missing: 61.4\%}
  & $<0.0001$ \\

Reaction rate (\%)
  & 4.1\%
  & 4.8\%
  & 1.9\%
  & 0.063 \\
\bottomrule
\end{tabularx}

\vspace{0.3ex}
{\footnotesize Continuous variables are mean (SD); categorical variables are n (\%).}
\end{table*}
\FloatBarrier

Table~\ref{tab:cluster-characteristics-K3} presents the key baseline characteristics for each of the three patient clusters. Cluster 1 is slightly older on average (55.4 ± 11.8) with a moderate number of cycles (5.1 ± 4.9). Cluster 2 is younger (51.8 ± 12.9) and receives the fewest cycles (4.0 ± 1.5). Cluster 3 is similarly young (51.5 ± 11.4) with the highest cycle count (5.6 ± 1.8) and the lowest reaction rate (1.9\%). Overall, clusters show distinct baseline profiles and treatment intensity (cycles), with notably low HSR incidence in Cluster 3.

P-values in the final column were calculated using Kruskal–Wallis tests for continuous variables (age and number of cycles) and chi-square tests for categorical variables (race, prior radiation, stage distribution, and reaction rate); Differences were observed for age ($p=0.001$), number of cycles ($p<0.0001$), prior radiation ($p<0.0001$), and {stage distribution} ($p<0.0001$); {race} did not differ across clusters ($p=0.603$), and the {reaction rate} difference did not reach the conventional significance threshold ($p=0.063$). 

\section*{Discussion}

This study was designed to answer a practical question faced by many researchers: when treatment-assignment heterogeneity is suspected but only partially reflected in observed baseline covariates, which propensity-score strategy should be used, and under what conditions does the added complexity of cluster-informed adjustment pay off? Our comparison across simulation scenarios and a real clinical application suggests a nuanced answer. Both cluster-informed strategies (fitting separate propensity-score models within empirically derived clusters, or including cluster membership as a covariate in a single global model) substantially reduced the bias and MSE introduced by omitted-covariate misspecification relative to standard IPW. In other words, accounting for empirical subgroup structure during propensity-score estimation improves robustness to a common and difficult-to-detect failure mode of IPW, regardless of which of the two implementations is chosen.

At the same time, neither cluster-informed strategy uniformly dominated the other, and the choice between them involves a genuine tradeoff. Clustering + IPW achieved lower MSE under correctly specified latent-cluster structure at every evaluated sample size, reflecting the benefit of allowing the full treatment-assignment relationship to vary across clusters. The global propensity-score model with cluster labels, in turn, generally achieved lower bias and better confidence-interval coverage, particularly at smaller sample sizes, likely because it estimates fewer parameters and is less vulnerable to small-cluster instability. For researchers, this suggests a simple rule: when the analytic priority is minimizing overall estimation error and sample sizes are adequate to support within-cluster modeling, Clustering + IPW is preferable; when the priority is preserving valid inferential coverage in smaller samples, or when simplicity and ease of implementation are valued, the global cluster-label model is a reasonable default. Regardless of the propensity-score strategy selected, diagnostics of covariate balance and treatment overlap remain essential, particularly because diagnostic assessment remains inconsistently reported in applied medical research \cite{Granger2020PSDiagnostics}.

Existing subgroup propensity-score methods, such as those of Yang et al. \cite{Yang2021CausalSubgroupPSW} and Dong et al. \cite{Dong2020SBPS}, generally assume subgroups are known or clinically prespecified in advance. Our comparison instead addresses the setting in which subgroup structure is not prespecified but is partially recoverable from baseline covariates using unsupervised clustering, which is a common situation in applied medical research where clinically meaningful heterogeneity is suspected but not formally defined. Within that setting, our contribution is the rigorous, sample-size- and misspecification-stratified comparison itself, together with concrete implementation guidance that we believe is directly usable by applied researchers facing this problem.

We also note two limitations. First, our simulations assumed a constant treatment effect across clusters; we did not evaluate how well either cluster-informed strategy recovers genuinely heterogeneous cluster-specific treatment effects, which is a natural direction for future work. Second, in the carboplatin application, only 29 hypersensitivity events were observed across the cohort, and the resulting cluster-specific odds ratios (Table~\ref{tab:cluster-specific-K3}) are best interpreted as exploratory descriptive evidence of possible treatment-effect variation across subgroups. Given the limited number of events in the present cohort, external validation in an independent dataset would meaningfully strengthen confidence in these cluster-specific patterns. We see this as an important direction for future work and are pursuing collaborative relationships with institutions including the University of Chicago and Rush University Medical Center to support such validation.

\backmatter




\section*{Statements and Declarations}

\bmhead{Code availability} R code, saved simulation results, knitted analysis outputs, and figures are provided in the Supplementary Material.
\bmhead{Funding} This work was supported in part by the National Cancer Institute Cancer Center Support Grant P30CA060553, the National Cancer Institute under R21CA277285 and R01CA298377, and by the Department of Defense under Award Number HT94252410972.
\bmhead{Competing interests}
The authors declare that they have no conflict of interest.
\bmhead{Author contribution} R.C. conceived the study, developed the methodology, supervised the project, and wrote the manuscript. S.Z. performed the simulations and statistical analyses, contributed to the methodology, and co-wrote the manuscript. W.S. and S.P. contributed to the clinical application and interpretation of the results. W.X., L.P., L.Z., and H.Z. contributed to the study design and revised the manuscript. All authors reviewed and approved the final manuscript.
\bibliography{ClusteringReference}

\end{document}